\documentclass[prb,twocolumn,showpacs,amsmath,amssymb,superscriptaddress]{revtex4}
\usepackage{graphicx}% Include figure files

\begin{document}
%\draft

\title{Optical properties of metallic (III,Mn)V ferromagnetic
semiconductors in the infrared to visible range}
\author{E. M. Hankiewicz}
\affiliation{Department of Physics, Texas A\&M University, College
Station, TX 77843-4242}
\author{T. Jungwirth}
\affiliation{Institute of Physics, ASCR, Cukrovarnick\'a 10, 162 53
Praha 6, Czech Republic}
\affiliation{Department of Physics, University of Texas, Austin,
TX 78712-0264}
\author{T. Dietl}
\affiliation{Institute of Physics, Polish Academy of Sciences, al.\
Lotnik\'ow 32/46, PL-02-668 Warszawa, Poland }
\affiliation{ERATO Semiconductor Spintronics Project, Japan Science and Technology Agency, al.\ Lotnik\'ow 32/46, PL-02-668 Warszawa, Poland}
\author{C. Timm}
\affiliation{Institut f\"ur Theoretische Physik, Freie Universit\"at
Berlin, Arnimallee 14, D-14195 Berlin, Germany}
\author{Jairo Sinova}
\affiliation{Department of Physics, Texas A\&M University, College
Station, TX 77843-4242}

\date{\today}

\begin{abstract}
We report on a study of the ac conductivity and magneto-optical
properties of metallic ferromagnetic (III,Mn)V semiconductors in
the infrared to visible spectrum at zero temperature. Our analysis is based on the
successful kinetic exchange model for (III,Mn)V ferromagnetic
semiconductors. We perform the calculations within the Kubo
formalism and treat the disorder effects pertubatively within the
Born approximation, valid for the metallic regime. We consider an
eight-band Kohn-Luttinger model (six valence bands plus two
conduction bands) as well as a ten-band model with additional
dispersionless bands simulating phenomenologically the
upper-mid-gap states induced by antisite and interstitial impurities.
These models qualitatively account for optical-absorption
experiments and predict new features in the mid-infrared Kerr
angle and magnetic-circular-dichroism properties as a function of
Mn concentration and free carrier density.
\end{abstract}


\maketitle
\section{INTRODUCTION}

The desire to integrate storage capabilities and information
processing in single semiconductor-based devices has fueled the
development of ferromagnetic semiconductors based on (III,Mn)V
materials.\cite{Ohno:1998_a,Ohno99,Dietl:2001_a,Dietl:2003_} These
materials have been the focus of intensive research over the
recent years after non-equilibrium growth procedures
\cite{Ohno:1992_,Ohno:1996_} have demonstrated the ability to
achieve a ferromagnetic phase by increasing the hole-carrier
concentration. Holes are introduced by Mn, which acts as an
acceptor when substituted for a cation. The improvement in the
carrier concentration is due to a decrease in the concentration of
compensating defects, such as As-antisites and Mn-interstitials.
These defects act as donors and hence reduce the free-carrier
concentration with respect to the substitutional-Mn density. The
donor defect concentration can be diminished even further by low
temperature annealing of these
materials.\cite{Hayashi:2001_,Edmonds:2002_b,Ku:2003_,Yu:2002_a,Edmonds:2004_,Kuryliszyn-Kudelska:2003_cond-mat/0304622}
Electron paramagnetic resonance (EPR) and optical measurements
\cite{Schneider:1987,Linnarsson:1997_,Szczytko:2001_,Szczytko:1999_b,Fedorych:2002}
show that the d-electrons of the Mn impurity are strongly
localized with total spin $S=5/2$. Also, the neutral impurity
state consisting of a 3d$^5$ configuration plus a weakly bound
hole is only observed experimentally for very low Mn
concentrations outside the doping range where ferromagnetism occurs.
Therefore, the holes are considered to reside in 
delocalized or weakly localized valence-band states.

For the doping regime where the highest conductivities are obtained, these
experimental considerations lead to a popular model starting from
a Hamiltonian of the form
\begin{eqnarray}
H&=&H_{\mathrm{host}}+J_{\mathrm{pd}}\sum_{I,i} \mathbf{S}_{I}\cdot \mathbf{s}_{i,\mathrm{holes}}
  \,\delta(\mathbf{r}_i-\mathbf{R}_I)\nonumber\\
&&+J_{\mathrm{sd}}\sum_{I,i} \mathbf{S}_{I}\cdot \mathbf{s}_{i,\mathrm{elec}}
  \,\delta(\mathbf{r}_i-\mathbf{R}_I),
\label{model}
\end{eqnarray}
where $H_{\mathrm{host}}$ is the Hamiltonian describing the electronic
structure of the host semiconductor, which can be limited to a given subset of the bands
such as the valence bands (holes), conduction bands (electrons), or both,
and the last two terms are the exchange coupling between the localized d-electrons and the
delocalized holes (p) and electrons (s).\cite{Dietl:2000_} This effective Hamiltonian
incorporates the effects of spin-orbit coupling,
which plays a very important role in the qualitative and quantitative
understanding of these materials.\cite{Dietl:2001_b,Kon02} The properties
predicted by this model are most easily understood in the strongly
metallic regime for which disorder in the spatial distribution of
the Mn$^{2+}$ ions and other  defects of the materials can  be
treated perturbatively. The model can be further simplified
by the mean-field approximation where usually the virtual
crystal approximation (VCA) is made.\cite{Dietl:2000_,Kon02} Disorder effects have been
taken into account by the
introduction of a finite quasiparticle lifetime within the Born
approximation. The accurate description of
many  thermodynamic  and   transport properties  of metallic
(Ga,Mn)As samples, such as the transition
temperature,\cite{Dietl:2000_,Dietl:2001_b,Jungwirth:2002_b} the anomalous Hall
effect,\cite{Jungwirth:2003_b,Jungwirth:2002_a} the anisotropic
magnetoresistance,\cite{Jungwirth:2003_b,Jungwirth:2002_c} the
magneto-crystalline anisotropy,\cite{Dietl:2001_b,Abolfath:2001_b,Konig:2001_b}
the spin-stiffness,\cite{Konig:2001_b}  the ferromagnetic domain
wall widths,\cite{Dietl:2001_c} the magnetic dynamic damping
coefficients,\cite{Sinova:2003_cond-mat/0308386} and the
magneto-optical
properties,\cite{Dietl:2001_b,Sinova:2002_,Sinova:2003_}  has
proven the merit of this effective Hamiltonian approach and the
perturbative treatment of disorder within the metallic regime. Disorder
effects in diluted magnetic semiconductors, including the case of strong
disorder, have recently been reviewed in Ref.~\onlinecite{Timm03_b}.

Optical properties  are among the key  probes into the electronic
structure of materials. Given the wide energy range that optical probes can
attain, many different physical effects can  be addressed by exploring the
full spectrum of  phenomena including the visible and infrared
absorption,
\cite{Singley:2002_,Singley:2003_,Chapman:1967_,Hirakawa:2002_,Hirakawa:2001_,Nagai:2001_,Szczytko:1999_a} magneto-optical
effects  such as the  magnetic circular
dichroism,\cite{Szczytko:1996_,Ando:1998_,Szczytko:1999_a,Beschoten:1999_}
Raman scattering,\cite{Sapega:2001_,Seong:2002_,Limmer:2002_}
photoemission,
\cite{Okabayashi:1998_,Okabayashi:1999_,Okabayashi:2002_,Okabayashi:2001_,Asklund:2002_},
cyclotron resonance,\cite{Khodaparast:2003_cond-mat/0307087,Sanders:2003_cond-mat/0304434,Mitsumori:2003_cond-mat/307268} and ellipsometry measurements

In earlier work some of us explored the theoretical predictions for the
infrared magneto-optical properties  of these materials within a six-band
model and found them to agree quantitatively with experiments in
the strongly metallic regime.\cite{Sinova:2002_,Yang:2003_,Sinova:2003_}
With better samples and a wider spectrum now available in experiments,
there is the need to explore the predictions in this broader range. Hence
in this paper we extend our model to the whole range from infrared to
optical frequencies. We calculate the infrared and optical conductivity, as
well as magneto-optical effects, for the typical level of compensation in
samples measured so far\cite{Singley:2002_,Singley:2003_,
Hirakawa:2002_,Hirakawa:2001_,Nagai:2001_} as well as for the more metallic
regime of optimally annealed samples, for which our theory is most suitable.
We also compute the dependence of magneto-optical effects on compensation  and Mn concentration, which
should be useful for detailed comparison with experiments,
providing further insights in the validity of this model.  The
strong effects predicted by our calculations should motivate
further experimental exploration of the complex magneto-optical
effects in these materials beyond the optical regime. We use the
$\mathbf{k}\cdot\mathbf{p}$ or Kohn-Luttinger (KL)
model\cite{Kohn:1957_} to describe the carriers, which are coupled
with the $S=5/2$ Mn moments by p-d exchange treated in the
mean-field approximation. \cite{Dietl:2000_} We consider an
eight-band (six valence plus two conduction bands) KL model as
well as a ten-band model with additional dispersionless bands
simulating the localized upper mid-gap states induced by the
antisite impurities, interstitial impurities, and possibly by 
intra-d-shell transitions within the Mn.\cite{Dietl:2001_b}

Dynamical-mean-field-theory studies,  \cite{Hwang:2002_} which use a
single-band model that neglects the  spin-orbit coupling and  the
heavy-light degeneracy  of a III-V   semiconductor  valence   band,
demonstrate  that   non-Drude impurity-band-related peaks  in the
frequency-dependent conductivity occur  in  DMS  ferromagnet  models  when
the  strength  of the exchange interaction is comparable  to the
valence-band  width. Although these insights are useful, the conductivities
predicted by this model are inconsistent with experiment  in their
temperature  dependence
%and  appear not  to be  in  the same regime as  experimental samples
and do not, by construction, incorporate strong magneto-optical effects in
the mid-infrared regime which we find to be important. In a recent related
investigation of the optical properties using Monte Carlo simulations, the
transition from the valence band to a Mn impurity band was studied.
\cite{Alvarez:2003_} However these approaches miss the important multi-band
structure of (III,Mn)V which seems to be a key to understand the
magneto-optical properties of these compounds in the metallic regime and
the anisotropies mentioned above.

The paper is organized as follows. In Sec.~\ref{theory} we briefly describe
our theoretical approach and approximations and present the model
Hamiltonian. In Secs.~\ref{8band} and \ref{10band} we analyze the results
for the eight- and ten-band models. We stress the strong dependence
between the concentration of carriers and the amplitude and shift of
magneto-optical effects. We compare theoretical results with experimental
ones emphasizing the role of the localized band in explaining optical
experiments.\cite{Singley:2003_} We summarize our results in Sec.~\ref{summary}.

\section{Theoretical approach}
\label{theory}

Our theoretical approach starts by coupling  the valence-band   electrons
with $S=5/2$ Mn local moments with  a semi-phenomenological local exchange
interaction, as shown in Eq.~(\ref{model}). We employ an eight- or ten-band
KL Hamiltonian $H_{\mathrm{KL}}$ for the carriers in the host GaAs and
treat the Mn local moments within mean-field theory and
VCA.\cite{Dietl:2000_} At zero temperature this  gives rise to valence and conduction 
bands  splitting  by  the   effective  exchange fields
$\mathbf{h}_{\mathrm{pd/sd}}=N_{\mathrm{Mn}}S J_{\mathrm{pd}/\mathrm{sd}}\hat{\mathbf{z}}$,   where
$N_{\mathrm{Mn}}$  is  the substitutional Mn density and the strength of
the exchange coupling between the valence (p) and conduction (s) electrons is taken to  be
$J_{\mathrm{pd}}=55\,\mathrm{meV}\,\mathrm{nm}^{-3}$ 
and $J_{\mathrm{pd}}=-9\,\mathrm{meV}\,\mathrm{nm}^{-3}$ respectively
as obtained from photoemission, resistance, and magnetic-circular-dichroism (MCD) measurements.
\cite{Okabayashi:1998_,Omiya:2000_,Szczytko:2001_a} We assume that the
magnetization is aligned along the growth ($\hat{\mathbf{z}}$) direction by
a  small external  magnetic field.   We also restrict  ourselves  to the
$T=0$ limit,  which allows us to neglect  scattering  off  thermal
fluctuations in the orientation of Mn moments. We assume collinear
magnetization  in  the  ground  state,  ignoring  the  possibility  of
disorder-induced  non-collinearity in the ground state  which is known to
be less likely for  the strongly metallic (III,Mn)V ferromagnets which we
focus on.\cite{Schliemann:2002_b,Zarand:2002_,Brey:2003_}

The ac-conductivity tensor is calculated within the Kubo formalism
described in earlier studies.\cite{Sinova:2002_,Sinova:2003_} The sources
of disorder known  to be  relevant in  these materials include positional
randomness  of the substitutional Mn ions with charge $Q=-e$ and random
placement of interstitial Mn ions and As antisites, both acting as
non-magnetic double donors with charge $Q=+2e$. \cite{Blinowski:2003_}
Previous estimates of the valence-band quasiparticle  lifetimes  using
Fermi's golden rule including screened Coulomb and exchange interactions
and the effect of compensation are of the order of 100 meV.
\cite{Jungwirth:2002_c} We introduce this disorder effect within the Born
approximation in the Kubo formalism as in
Refs.~\onlinecite{Sinova:2002_} and \onlinecite{Sinova:2003_}.

In the calculations the Mn concentration $x$ and the free-carrier
concentration $p$ are treated as independent parameters. Compensation of
the holes introduced by the Mn acceptors is due to
antisite and interstitial defects,
%present due to the non-equilibrium growth techniques
as noted above. Recently developed annealing procedures allow
some independent experimental control over $x$ and $p$.\cite{Ohno99,%
Edmonds:2002_b,Ku:2003_,Mathieu:2002_cond-mat/0208411,Yu:2002_a,%
Edmonds:2004_,Kuryliszyn-Kudelska:2003_cond-mat/0304622}

\subsection{Model Hamiltonians}
\label{KL}

In the VCA, the interactions are replaced by
their spatial  averages, so that the Coulomb interaction vanishes and
holes  interact with a homogeneous exchange  field. The  unperturbed
Hamiltonian  for  the valence and conduction bands reads
$H_0=H_{\mathrm{KL}}+\mathbf{h} \cdot\mathbf{s}$, where $H_{\mathrm{KL}}$
is the eight-band KL Hamiltonian of pure GaAs, and $\mathbf{h}$ is the
exchange field that splits the valence and conduction bands respectively
as mentioned in the previous section. To take into account the
considerable contribution from donor defect-induced states below the
conduction-band edge, we phenomenologically add two dispersionless bands.
We assume that the latter are composed of localized states of mainly
s-type. Choosing the  angular momentum quantization direction  along
the $z$-axis, the basis functions corresponding to heavy, light, split-off
holes, conduction-band electrons, and defect bands can be written, in this
order, as
\begin{eqnarray*}
|1\rangle & = & \frac{-1}{\sqrt{2}}(X_{\uparrow}+iY_{\uparrow}) ,\\
|2\rangle & = & \frac{1}{\sqrt{6}}(X_{\uparrow}-iY_{\uparrow})+
\sqrt{\frac{2}{3}}Z_{\downarrow} ,\\
|3\rangle & = & \frac{-1}{\sqrt{6}}(X_{\downarrow}+iY_{\downarrow})+
\sqrt{\frac{2}{3}}Z_{\uparrow} ,
\end{eqnarray*}

\begin{eqnarray*}
|4\rangle & = & \frac{1}{\sqrt{2}}(X_{\downarrow}-iY_{\downarrow}) ,\\
|5\rangle & = & \frac{-1}{\sqrt{3}}(X_{\downarrow}+iY_{\downarrow})-
\sqrt{\frac{1}{3}}Z_{\uparrow} ,\\
|6\rangle & = & \frac{-1}{\sqrt{3}}(X_{\uparrow}-iY_{\uparrow})+
\sqrt{\frac{1}{3}}Z_{\downarrow} ,\\
\end{eqnarray*}

\begin{eqnarray*}
|7\rangle & = & S_{\uparrow} ,\\
|8\rangle & = & S_{\downarrow} ,\\
|9\rangle & = & S^{\mathrm{an}}_{\uparrow} ,\\
|10\rangle & = & S^{\mathrm{an}}_{\downarrow} .
\end{eqnarray*}
where $X$,$Y$,$Z$ are p-like orbitals, $S$ and $S_{an}$ are the s
orbitals associated with conduction and defect bands respectively.
In this basis, the ten-band KL Hamiltonian has a form
\begin{widetext}
\begin{equation}
H_{\mathrm{KL}} = \left(
\begin{array}{cccccccccc}
  H_{\mathrm{hh}} & -c& -b & 0 & \frac{b}{\sqrt{2}} & c\sqrt{2} &e &
  \multicolumn{1}{c|}{0} &  e_{\mathrm{an}}  &  0  \\
  -c^* & H_{\mathrm{lh}}& 0 & b & -\frac{b^{*}\sqrt{3}}{\sqrt{2}} & -d &
  \frac{1}{\sqrt{3}}e^{*} & \multicolumn{1}{c|}{f^{*}} &
  \frac{1}{\sqrt{3}}e_{\mathrm{an}}^{*}  & f_{\mathrm{an}}^{*}  \\
  -b^* & 0 & H_{\mathrm{lh}}& -c & d &  -\frac{b\sqrt{3}}{\sqrt{2}}  &
  f^{*} &\multicolumn{1}{c|}{\frac{e}{\sqrt{3}}} & f_{\mathrm{an}}^{*} &
  \frac{e_{\mathrm{an}}}{\sqrt{3}}\\
  0  & b^*& -c^* & H_{\mathrm{hh}} & -c^*\sqrt{2} & \frac{b^*}{\sqrt{2}} &
  0 & \multicolumn{1}{c|}{e^{*}} &  0  & e_{\mathrm{an}}^{*}\\
  \frac{b^*}{\sqrt{2}} & -\frac{b\sqrt{3}}{\sqrt{2}}& d^* & -c\sqrt{2} &
  H_{\mathrm{so}} & 0 & \frac{f}{\sqrt{2}} &
  \multicolumn{1}{c|}{\sqrt{\frac{2}{3}}e} &
  \frac{f_{\mathrm{an}}}{\sqrt{2}} & \sqrt{\frac{2}{3}}e_{\mathrm{an}} \\
  c^*{\sqrt{2}}& -d^*& -\frac{b^{*}\sqrt{3}}{\sqrt{2}}  &
  \frac{b}{\sqrt{2}}  & 0 & H_{\mathrm{so}} & -\sqrt{\frac{2}{3}}e^{*} &
  \multicolumn{1}{c|}{\frac{f^{*}}{\sqrt{2}}}  &
  -\sqrt{\frac{2}{3}}e_{\mathrm{an}}^{*} &
  \frac{f_{\mathrm{an}}^{*}}{\sqrt{2}} \\
  e^* & \frac{1}{\sqrt{3}}e  & f &  0  & \frac{f^*}{\sqrt{2}} &
  -\sqrt{\frac{2}{3}}e & H_{\mathrm{cb}} & \multicolumn{1}{c|}{0} & 0 & 0\\
  0  &  f  & \frac{e^{*}}{\sqrt{3}} &  e  & \sqrt{\frac{2}{3}}e^* &
  \frac{f}{\sqrt{2}} & 0 & \multicolumn{1}{c|}{H_{\mathrm{cb}}} & 0 & 0 \\
\cline{1-8}
  e_{\mathrm{an}}^{*} & \frac{1}{\sqrt{3}}e_{\mathrm{an}}  & f_{\mathrm{an}} &  0  & \frac{f_{\mathrm{an}}^{*}}{\sqrt{2}} & -\sqrt{\frac{2}{3}}e_{\mathrm{an}} & 0 & 0 & E'_{g} & 0 \\
   0  & f_{\mathrm{an}} & \frac{e_{\mathrm{an}}^{*}}{\sqrt{3}} &  e_{\mathrm{an}}  & \sqrt{\frac{2}{3}}e_{\mathrm{an}}^{*} & \frac{f_{\mathrm{an}}}{\sqrt{2}} & 0 & 0 & 0 & E'_{g} \\
\end{array}
\right) ,
\end{equation}
\end{widetext}
where the eight-band Hamiltonian is the highlighted sector. The
quantities that appear in $H_{KL}$  are:
\begin{eqnarray}
H_{\mathrm{cb}} & = & E_g +
\frac{\hbar^2}{2m^{*}_c}\,(k_x^2+k_y^2+k_z^2); \\
H_{\mathrm{hh}} &
= & -\frac{\hbar^2}{2m_{0}}\,
  \bigg[\bigg(\gamma_{1}+\gamma_{2}-\frac{E_{\mathrm{an}}}
  {2(E'_g+\Delta_{\mathrm{so}}/3)} \nonumber\\
& -&\frac{E_{P}}{2(E_g+\Delta_{\mathrm{so}}/3)}\bigg)(k_x^2+k_y^2)
  +(\gamma_{1}-2\gamma_2)k_z^2\bigg];  
\end{eqnarray} %\\
\begin{eqnarray}
H_{\mathrm{lh}} & = & -\frac{\hbar^2}{2m_{0}}\,
  \bigg[\bigg(\gamma_{1}-\gamma_{2}-\frac{E_{\mathrm{an}}}
  {6(E'_g+\Delta_{\mathrm{so}}/3)} \nonumber\\
&&
{}-\frac{E_{P}}{6(E_g+\Delta_{\mathrm{so}}/3)}\bigg)\,(k_x^2+k_y^2)
  +\bigg(\gamma_{1}+2\gamma_2 \nonumber \\
&& {}-\frac{E_{\mathrm{an}}}{3(E'_g+\Delta_{\mathrm{so}}/3)}
  -\frac{E_{P}}{3(E_g+\Delta_{\mathrm{so}}/3)}\bigg)k_z^2\bigg];
\end{eqnarray}
\begin{eqnarray}
H_{\mathrm{so}} & = &-\Delta_{\mathrm{so}} -\frac{\hbar^2}{2m_0}
\bigg(\gamma_{1}-\frac{E_{\mathrm{an}}}
  {3(E'_g+\Delta_{\mathrm{so}}/3)}\nonumber\\
 && {} -\frac{E_{P}}{3(E_g+\Delta_{\mathrm{so}}/3)}\bigg)
(k^2_{x}+k^2_{y}+k^2_{z});\\
b & = & -\frac{\hbar^2}{2m_0}\,\sqrt{12}\,
  \bigg(\gamma_{3}-\frac{E_{\mathrm{an}}}
  {6(E'_g+\Delta_{\mathrm{so}}/3)} \nonumber\\
&& {}-\frac{E_{P}}{6(E_g+\Delta_{\mathrm{so}}/3)}\bigg)\,
  k_{z}\,(k_x-ik_{y}) 
\end{eqnarray}
\begin{eqnarray}
c & = & -\frac{\hbar^2}{2m_0}\,\sqrt{3}\,\bigg[\bigg(
  \gamma_{2}-\frac{E_{\mathrm{an}}}
  {6(E'_g+\Delta_{\mathrm{so}}/3)} \nonumber\\
&& {}-\frac{E_{P}}{6(E_g+\Delta_{\mathrm{so}}/3)}\bigg)\,
  (k_{x}^{2}-k_{y}^{2}) \nonumber\\
&& {}-2i\,\bigg(\gamma_{3}-\frac{E_{\mathrm{an}}}
  {6(E'_g+\Delta_{\mathrm{so}}/3)}\nonumber\\
&& {}  -\frac{E_{P}}{6(E_g+\Delta_{\mathrm{so}}/3)}\bigg)k_xk_y \bigg];
\end{eqnarray}
\begin{eqnarray}
d & = &
-\frac{\hbar^2}{2m_0}\,\sqrt{2}\,\bigg[-\bigg(2\gamma_{2}
  -\frac{E_{\mathrm{an}}}{3(E'_g+\Delta_{\mathrm{so}}/3)} \nonumber\\
&& {}-\frac{E_{P}}{3(E_g+\Delta_{\mathrm{so}}/3)}\bigg)\,k_{z}^{2}
  +\bigg(\gamma_{2}-\frac{E_{\mathrm{an}}}
  {6(E'_g+\Delta_{\mathrm{so}}/3)}\nonumber\\ &&
{}-\frac{E_{P}}{6(E_g+\Delta_{\mathrm{so}}/3)}\bigg)\bigg]\,(k_x^2+k_y^2),
\end{eqnarray}

\begin{equation}
e = \frac{iP}{\sqrt{2}}(k_x-ik_y) , \quad f = \sqrt{\frac{2}{3}}iP
k_z ,
\end{equation}
\begin{equation}
e_{\mathrm{an}} = \frac{iP_{\mathrm{an}}}{\sqrt{2}}(k_x-ik_y) , \quad f_{\mathrm{an}} =
\sqrt{\frac{2}{3}}iP_{\mathrm{an}} k_z ,
\end{equation}
where $E_g$ is the energy gap between valence and conducting
bands, $E'_g$ is the energy of the dispersionless defect band,
$m_0$ is the bare electron mass, $m^{*}_c$ is the effective mass
of conduction electrons, for GaAs $m^{*}_c=0.067 m_0$,
and $\Delta_{\mathrm{so}}$ is the energy of spin-orbit splitting. $P$ is the
momentum matrix element between conducting and valence bands
given by,
\begin{equation}
P = -\frac{i\hbar}{m_0}\, \langle
  S|\hat{p}_x|X\rangle .
\label{P}
\end{equation}
We introduce the dipole energy $E_P=2m_0 P^2/\hbar^2$, and
set $E_P=22.5\,\mathrm{eV}$ in accordance with
Ref.~\onlinecite{Ostromek:1996}. Similarly,
$P_{\mathrm{an}}$ is the momentum matrix
elements between defect level and valence bands defined as
\begin{equation}
P_{\mathrm{an}} = -\frac{i\hbar}{m_0}\, \langle
  S_{\mathrm{an}}|\hat{p}_x|X\rangle ,
\label{Pan}
\end{equation}
The corresponding energy associated with the defect level is
$E_{\mathrm{an}}=2m_0 P_{\mathrm{an}}^2/\hbar^2$. This energy is a parameter of our model. We
assume that $E_{\mathrm{an}}$ is of the order of few eV.

In the
six-band KL model the valence-band structure is parametrized by the Luttinger
parameters $\gamma_1$, $\gamma_2$, $\gamma_3$, and
$\Delta_{\mathrm{so}}$.
For GaAs the values are $\gamma_{1}=6.98$, $\gamma_2=2.06$,
$\gamma_3=2.93$, and $\Delta_{\mathrm{so}} = 341\,\mathrm{meV}$.
For the eight- and ten-bands KL model the corrections from the
conduction and defect bands must be consistently taken into
account in order to conserve spectral weight.

It has been shown that the energy gap in heavily doped GaAs
strongly depends on the carrier concentration.\cite{Casey75} There are
several effects which can cause the gap to be reduced, like
band-gap narrowing by many-body effects as well as the band-tailing\cite{Casey75,Mahan}

The energy gap $E_g$ is estimated in accordance with Ref.~\onlinecite{Casey75} by the
formula: $E_g(\mathrm{eV})=1.52-1.6\cdot
10^{-8}[\mathrm{p(cm^{-3})}]^{1/3}$. For simplicity, the disorder-induced
broadening, or quasiparticle lifetime, is taken to be a constant of the
order of $100$ meV in accordance with earlier
estimates.\cite{Jungwirth:2002_c}

\section{Results and discussion}
\subsection{Eight-band model}
\label{8band}

Since the infrared conductivity was investigated in detail in the
framework of the six-band model,\cite{Sinova:2002_} we first compare the
results of the two models in the expanded range.

\begin{figure}[tbh]
\includegraphics[width=3.3in,clip]{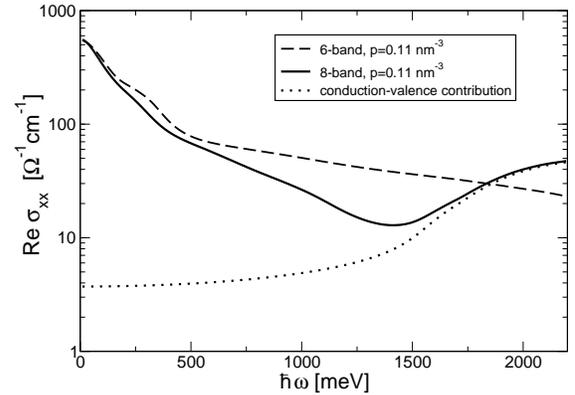}
\caption{ Real part of the diagonal conductivity,
$\mathrm{Re}\,\sigma_{xx}$, as a function of frequency $\hbar\omega$.
The dashed and solid lines correspond to the total
$\mathrm{Re}\,\sigma_{xx}$ for the six- and
eight-band models, respectively. The dotted line shows
$\mathrm{Re}\,\sigma_{xx}$ for \emph{only} valence-conduction-band
transitions within the eight-band model. The Mn concentration is 5\%, the
hole concentration is $p=0.11\,{\rm nm}^{-3}$, and the lifetime
broadening is $\Gamma=100\,\mathrm{meV}$.}
\label{fig1}
\end{figure}

In Fig.~\ref{fig1} the real part of the diagonal conductivity
$\sigma_{xx}(\omega)$ is plotted for the six- and eight-band models as a
function of frequency for a Mn concentration of 5\%, a hole concentration
of $p=0.11\,{\rm nm}^{-3}$, and a lifetime broadening of $\Gamma
=100\,\mathrm{meV}$. As discussed earlier,\cite{Sinova:2002_} the features
at low frequencies together with the shoulder around $220\,\mathrm{meV}$
correspond to the Drude peak combined with transitions between heavy- and
light-hole bands. The transition from the heavy-hole to the split-off band
around $500\,\mathrm{meV}$ and from the light-hole to the split-off band
around $900\,\mathrm{meV}$ are strongly broadened and therefore
unobservable.\cite{Sinova:2002_,Yang:2003_} For frequencies above
$500\,\mathrm{meV}$ one can see the loss of spectral weight at high
infrared frequencies in the eight-band model in comparison with the
six-band model, which is not considered reliable in this high-frequency
range. Of course the spectral weight is not lost but shifted to higher
photon energies above $1.5\,\mathrm{eV}$ as expected from the $f$-sum rule
where total integrated conductivity should remain constant. We identify the
upturn around $1.5\,\mathrm{eV}$ in the eight-band model as the
valence-conduction-band transition (see dotted line in Fig.~\ref{fig1}).

\begin{figure}[bt]
\includegraphics[width=3.3in,clip]{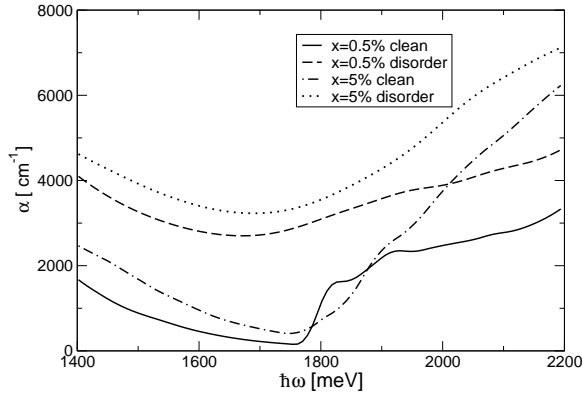}
\caption{Absorption as a function of photon energy for Mn
concentrations $x=0.5$\% and $x=5$\% and
$p=0.35\,\mathrm{nm}^{-3}$ for a clean system and for a
system with disorder induced by impurity scattering.}
\label{fig2}
\end{figure}

In Fig.~\ref{fig2} we show the absorption as a function of frequency for
clean and disordered systems with different concentrations of Mn. As a
check we first consider a low concentration of Mn, $x=0.5$\%. In the clean
limit we observe the expected sharp band edge at optical frequencies. When
including the disorder effect perturbatively the band edge gets broadened but still
identifiable within a few percent of the the result for the clean calculation.
For the typical concentration $x=5$\%
for ferromagnetic (Ga,Mn)As the absorption is stronger and the band edge also
broadened. Note also that due to the $f$-sum rule the spectral weight
associated with inter- and intra-valence-band transitions is shifted to
higher energies in the disordered case, as observed experimentally.

\begin{figure}[tbh]
\includegraphics[width=3.3in,clip]{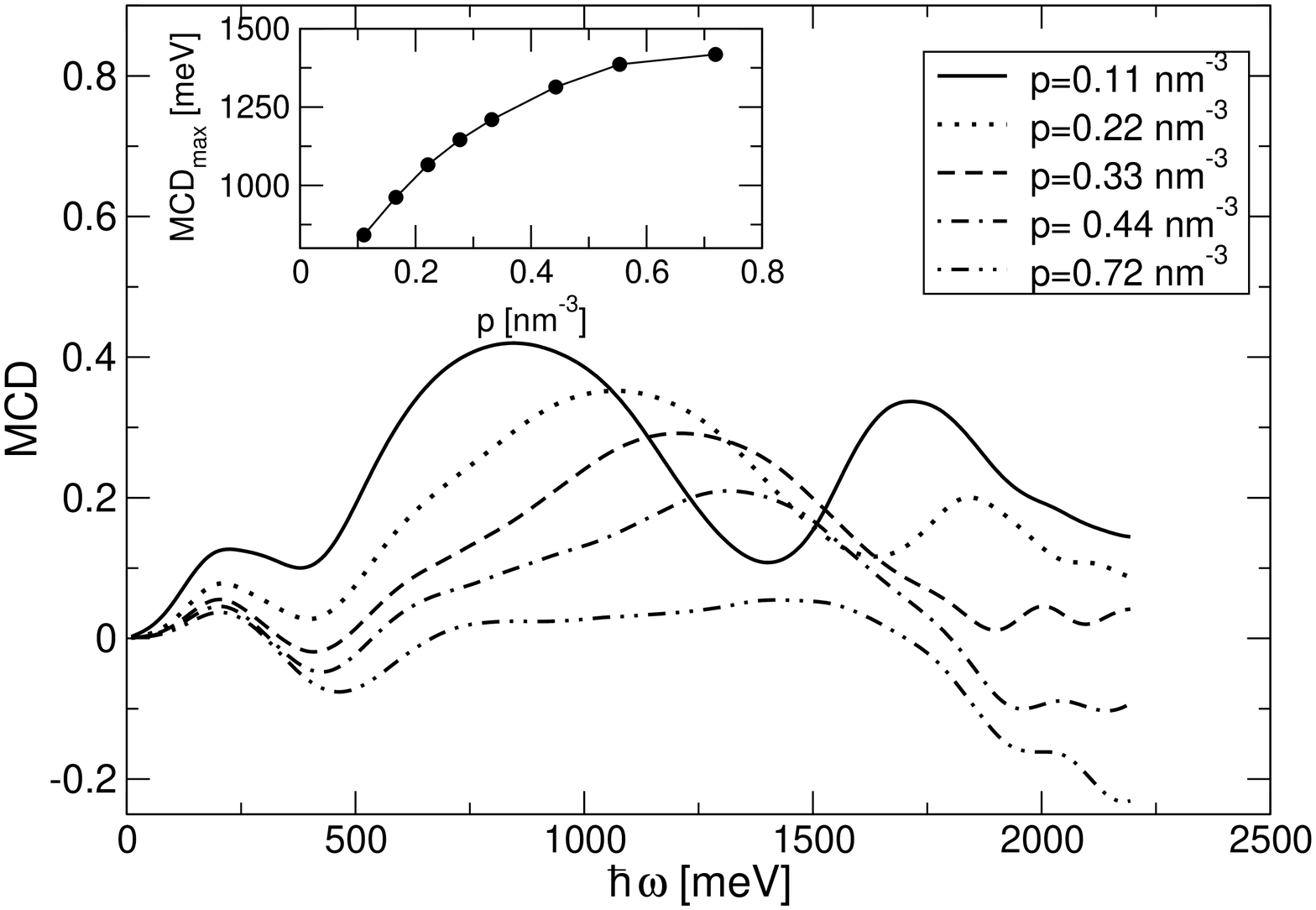}
\caption{Magnetic circular dichroism (MCD) as a function of
photon energy for different concentrations of holes for
Ga$_{0.95}$Mn$_{0.05}$As. The inset shows the position of the
maximum of the MCD spectrum \textit{vs.}\ hole concentration. The lifetime
broadening is $\Gamma=100\,\mathrm{meV}$.}
\label{fig3}
\end{figure}

{}From the point of view of integration of (Ga,Mn)As in magnetorecording
devices the strength of magneto-optical effects is very
important.\cite{Ando00} Therefore, we now present our predictions for
magnetic circular dichroism (MCD) as well as the Kerr effect as a function
of hole concentration. We consider the system air/(Ga,Mn)As/GaAs, taking
into account multiple reflections in the thin-film limit of (Ga,Mn)As,
\textit{i.e.}, for  wavelengths larger than the epilayer thickness.  We
postpone the discussion of magneto-optical effects in thicker films of
(Ga,Mn)As to the next subsection in the context of the  ten-band model.

The MCD signal is defined as the difference of absorption for right- and
left-circularly-polarized light renormalized by the total absorption.
In the thin-film geometry one obtains for the MCD signal
\begin{equation}
\mathrm{MCD} \equiv \frac{\alpha^{+}-\alpha^{-}}{\alpha^{+}+\alpha^{-}}
=\frac{\mathrm{Im}\,\sigma_{xy}(\omega)}{\mathrm{Re}\,\sigma_{xx}(\omega)} .
\label{mcd}
\end{equation}
In Fig.~\ref{fig3} we show the MCD signal as a function of frequency for
different concentrations of carriers. The strong p-d exchange in
diluted magnetic semiconductors causes the splitting of bands.
However, it is the strong spin-orbit coupling present in these
materials that gives rise to a strong MCD signal in the
mid-infrared range. This spin-orbit coupling and the exchange field
allow transitions between spin-orbit-coupled bands
which give rise to a high $\sigma_{xy}(\omega)$ in the mid-infrared regime.
The natural energy scales of the Fermi energy and the valence-band
splittings are in the infrared range and thus the difference in
absorption of differently polarized light should be particularly
pronounced in this range as is evident from the theoretical
curves. The maximum for small photon energies around $220\,\mathrm{meV}$
corresponds to heavy-hole to light-hole transitions while the main
maximum in the $800-1400 {\rm meV}$  of transitions between spin-orbit split-off holes and
heavy/light holes and between valence bands and the
conduction band.
The frequency of the main maximum is shown in
the inset as a function of hole concentration. Its position monotonically
increases with carrier concentration. This dependence of the
MCD signal could be useful for the estimation of carrier concentration
in experimentally measured (Ga,Mn)As thin films and should be a
straightforward test of the validity of this theory.

\begin{figure}[tbh]
\includegraphics[width=3.3in,clip]{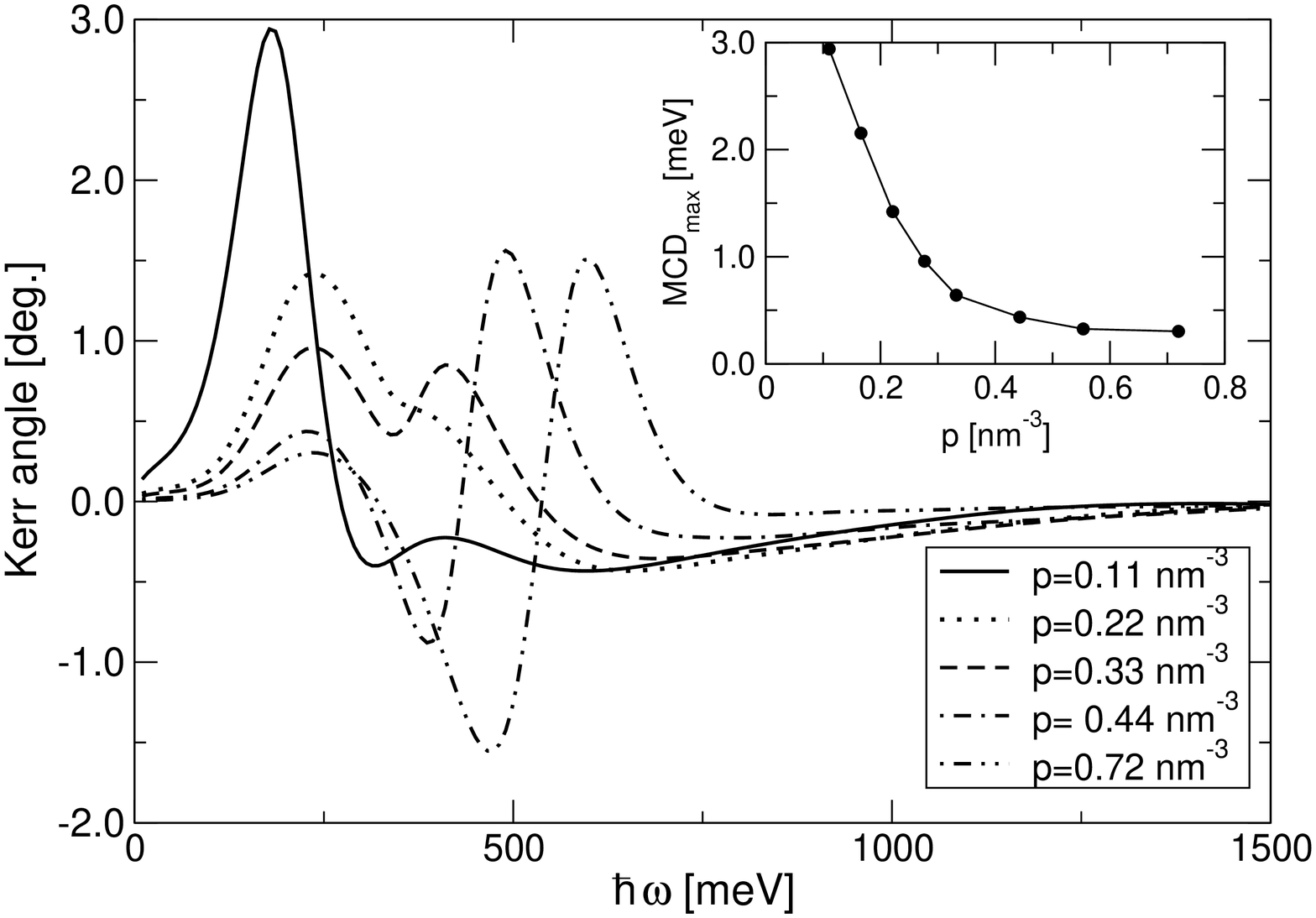}
\caption{Kerr angle $\Theta_{\mathrm{Kerr}}$
as a function of frequency for different
concentrations of holes for Ga$_{0.95}$Mn$_{0.05}$As. In the
inset we plot the amplitude of the Kerr angle at
$\hbar\omega=220\,\mathrm{meV}$ as a function of hole concentration. The
lifetime broadening is $\Gamma=100\,\mathrm{meV}$.}
\label{fig4}
\end{figure}

We next turn to the calculation of the Kerr effect,
which occurs in reflection from a magnetic medium. In the
thin-film geometry the Kerr angle $\Theta_{\mathrm{\mathrm{Kerr}}}$
and the ellipticity $\eta_{\mathrm{Kerr}}$ are defined as:
\cite{Ando00}
\begin{equation}
\label{Kerr}
\Theta_{\mathrm{Kerr}}+i\eta_{\mathrm{Kerr}} =\frac{r_{+}-r_{-}}{r_{+}+r_{-}}
\end{equation}
where $r_{\pm}$ are the total complex reflection amplitudes
for left- and right-circularly-polarized light. The
Kerr angle is shown in Fig.~\ref{fig4} as a function of the photon
energy for different carrier concentrations and a Mn
concentration of $x=5$\%. The Kerr angle is of the order of a few
degrees, comparable to the Kerr effect in materials used in
magnetorecording devices.\cite{Kaneko} In the inset, the value of
the Kerr angle at $220\,\mathrm{meV}$ is plotted as a function of hole
concentration. In contrast to the MCD signal one observes a
decrease of the Kerr angle around $220\,\mathrm{meV}$ with
increasing hole concentration.
The sharp changes of sign are again
associated with different transitions within the
valence subbands and the strong spin-orbit coupling.

\subsection{Ten-band model}
\label{10band}

{}From earlier studies on low-temperature GaAs (LT-GaAs) it is known that
defects affect the electrical and optical properties since they interact
with carriers by acting as traps or scattering and recombination
centers.\cite{Casey75,Weber82,Meyer84,Kauf88,Liu97,Kam85} Experiments
revealed the presence of arsenic antisites, arsenic interstitials, and
their complexes.\cite{Bourg88} Usually a Coulomb
potential is introduced to describe  effects of these charged
defects. The early papers studied the dependence of the energy gap on
the concentration of holes in heavily doped GaAs, using a one-band model
and a screened impurity Coulomb potential.\cite{Casey75} EPR measurements
in GaAs showed the formation of two antisites levels $0.52\,\mathrm{eV}$ and
$0.75\,\mathrm{eV}$ above the valence band.\cite{Weber82} However, the
interpretation of the MCD spectra has been controversial until recently:
According to Meyer \textit{et al.},\cite{Meyer84} MCD spectra arise from
intra-center transitions from the $\mathrm{A}_1$ s-type ground state of the
defect to two $\mathrm{T}_2$ excited states with p-character. On the other
hand, the model proposed by Kaufmann and Windscheif\cite{Kauf88} postulates
that MCD is a result of transitions between the s-state $\mathrm{A}_1$ of
the defect and the valence bands. Recent studies of arsenic antisites in
MCD support the second model.\cite{Liu97} The intensity of the MCD is very
weak in undoped GaAs, of the order of $5\cdot 10^{-3}$.

The situation becomes more complicated if GaAs is doped with Mn. During
low-temperature growth under arsenic overpressure the high concentration of
Mn leads to the formation of antisites and Mn interstitials, according to
\textit{ab-initio} calculations.\cite{Petuk02,Masek03}
Interstitials have been observed
in  channeling Rutherford backscattering experiments by Yu \textit{et
al}.\cite{Yu:2002_a} For a total manganese concentration of $x\approx 7$\%
about 17\% of the manganese impurities were found in interstitial
positions.\cite{Yu:2002_a} Theoretical studies predict that interstitials
form a localized level $0.9\,\mathrm{eV}$ above the valence
band.\cite{Petuk02} As an additional complication, a correlated spatial
distribution of defects is expected to develop during growth or annealing
due to the strong Coulomb interactions between the
defects.\cite{Timm02,Timm03_a,Timm03_b}
Singley \textit{et al.}\cite{Singley:2003_} emphasize that the observed
band-tail signal near $0.7\,E_g$ is likely associated with
defect-induced upper-mid-gap states as we consider here. 
However, recent experiments \cite{Wolos:preprint,Heitz:1997,Korotkov,Graf} suggest
that another candidate may be intra-d-shell transitions within the Mn
as predicted earlier by in Ref. \onlinecite{Dietl:2001_b}.
Hence, without specifying the nature of the mid-gap state we introduce
a phenomenological dispersionless band in the gap near $0.7\,E_g$
to model the system.

\begin{figure}[tbh]
\includegraphics[width=3.3in,clip]{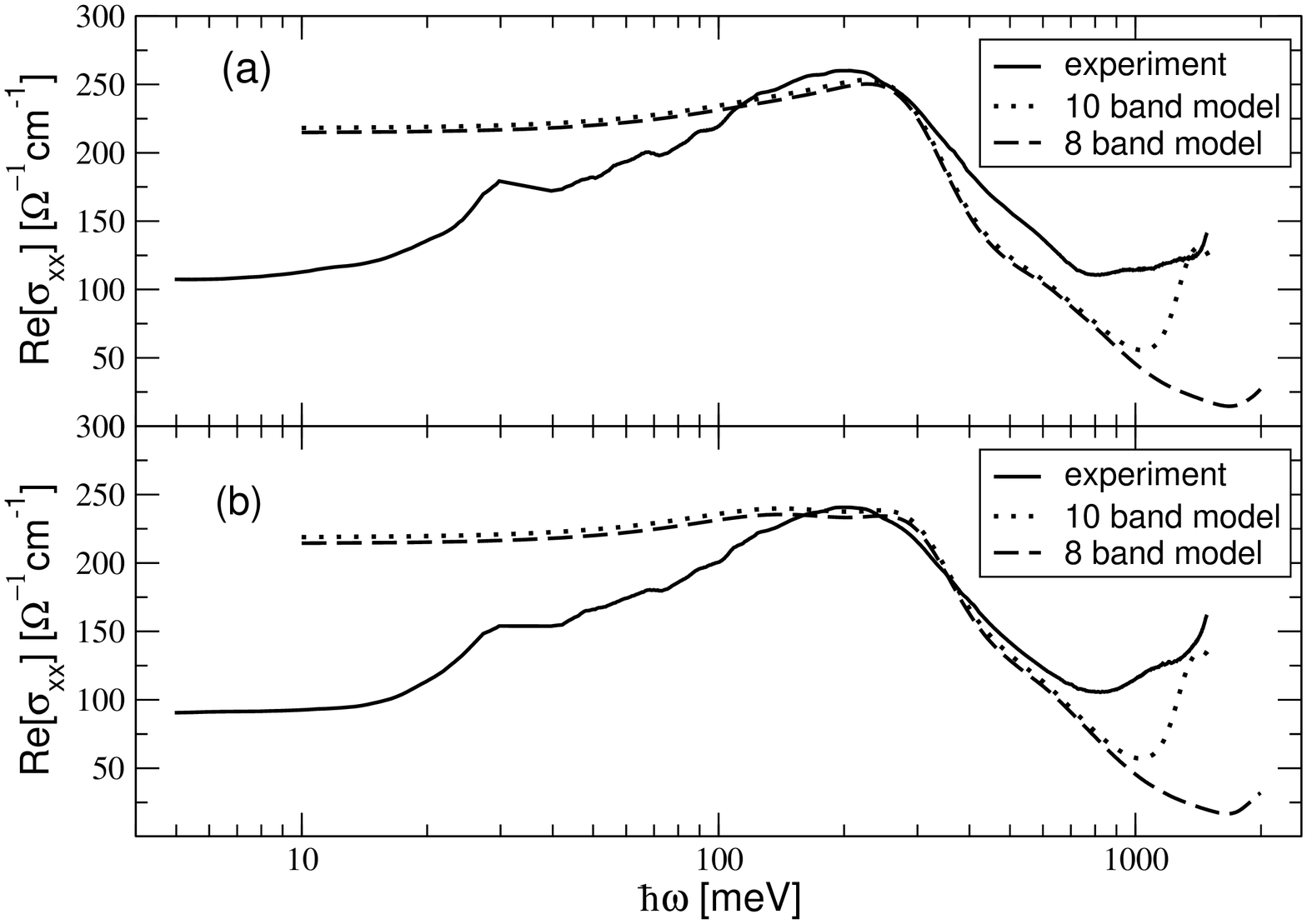}
\caption{Real part of the diagonal conductivity,
Re$[\sigma_{xx}(\omega)]$, as a function of frequency for (a)
x=5.2\% and (b) x=6.66\%. Here  we compare experimental results
(solid line) with theoretical model for  eight-bands (dotted line) and
ten-bands (dashed line) model calculate it on. The dispersionless
defect energy level,$E'_g$, is 1eV above the valence band.}
\label{fig5}
\end{figure}

In Fig.~\ref{fig5} we show the real part of the diagonal conductivity,
$\mathrm{Re}\,\sigma_{xx}$, as a function of frequency for the ten-band
model with Mn concentration $x=5.2$\% and $x=6.66$\% and carrier
concentration $p=0.345\,\mathrm{nm}^{-3}$ and $p=0.368\,\mathrm{nm}^{-3}$,
respectively, and a quasiparticle lifetime of $100\,\mathrm{meV}$.
Experimental data by Singley \textit{et al.}\cite{Singley:2003_} and theoretical
fits in the framework of the eight- and ten-band models are shown. The
intra-valence-band contributions are not included in the spectra shown because
the low-frequency conductivity is very small for these samples, of the
order of $100\,\Omega^{-1}\mathrm{cm}^{-1}$.
Exact diagonalization studies for these models show a strong suppression of the Drude peak (intraband
contribution) for ``weakly'' metallic samples due to partial localization
which cannot be captured by our model and hence its suppression from the
spectra shown in Fig.~\ref{fig5}.\cite{Yang:2003_}
From this theoretical comparison we estimate a
concentration of carriers of $p=0.35\,{\rm nm}^{-3}$ and $p=0.37\,{\rm
nm}^{-3}$ for Mn concentration of 5.2\% and 6.66\%, respectively, for the
samples measured by Singley {\it et al}.\cite{Singley:2003_} Moreover we find on
the basis of similar fits that the compensation in as-grown samples is
around 70--80\%.

The peak around $220\,\mathrm{meV}$ in  optical absorption experiments and
theoretical data is associated  with the inter-valence-subband transitions.
One can see that the experimental and theoretical positions of this
transition are in agreement.\cite{Singley:2002_,Singley:2003_}
We can identify the transitions between light-hole band and split-off band around
$500\,\mathrm{meV}$ in both experimental and theoretical data.
For frequencies larger than $700\,\mathrm{meV}$ the eight-band result does not
match the experimental data in the upturn observed experimentally around $1.5\,\mathrm{eV}$.
Note that within the ten-band model the upturn at high energies of
the order of $1.5\,\mathrm{eV}$ is reproduced. The experimental
EPR data for undoped GaAs show that antisite levels lie
$0.52\,\mathrm{eV}$ and $0.75\,\mathrm{eV}$ above the valence
band.\cite{Weber82} On the other hand, \textit{ab-initio} theory
\cite{Petuk02,Masek03} predicts interstial levels
$0.9\,\mathrm{eV}$ above the valence band. Comparison of our
numerical calculations  with experimental data\cite{Singley:2003_} could
thus suggest that the localized states near the conduction band
arise from interstitials rather than from antisites. However, the
discrepancy of theoretical and experimental fits near
$1\,\mathrm{eV}$ could also suggest a scenario with a broader
impurity band or a merging of the conduction band with the
impurity band.
Another possible scenario, suggested by photoluminescence and optical
absorption experiments on GaN:Mn,Mg
\cite{Wolos:preprint,Korotkov,Graf} and GaN:Fe \cite{Heitz:1997}, is that part of the
upturn in the spectra in the optical range is due to intra d-shell transitions.
We believe the resolution of this question can be determined by further
experiments on optimally annealed samples, for which this theory
is designed, in particular on metallic samples with dc
conductivities above a few hundred $\Omega^{-1}\mathrm{cm}^{-1}$.

\begin{figure}[tbh]
\includegraphics[width=3.3in,clip]{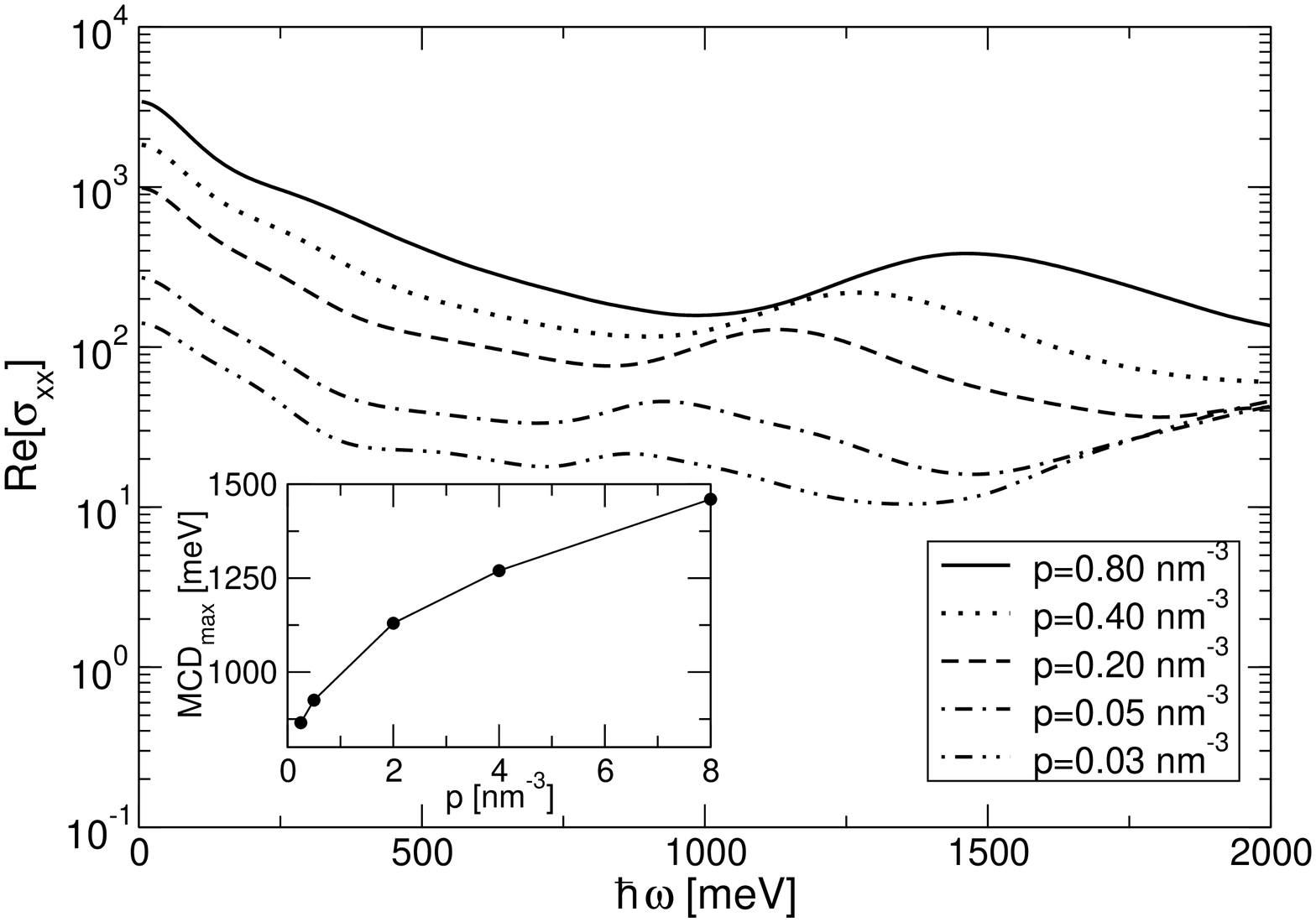}
\caption{Real part of the diagonal conductivity as a
function of frequency for Ga$_{0.96}$Mn$_{0.04}$As and various
hole concentrations. In the inset the position of the maximum
corresponding to transitions between valence and defect band is plotted
\textit{vs.}\
carrier concentration. Here, the dispersionless defect energy
level $E'_g$ is $0.75\,\mathrm{eV}$ above the valence band.
The dipole energy between the defect level and the valence band is
$E_{\mathrm{an}}=2.2\,\mathrm{eV}$.}
\label{fig6}
\end{figure}

In Fig.~\ref{fig6} the total real part of the diagonal conductivity is
plotted as a function of frequency for different concentrations of holes.
We assume that the localized states form a flat band $0.75\,\mathrm{eV}$
above the valence band.

 Thus on the basis of experimental data\cite{Weber82} for undoped GaAs they
might correspond to antisite states. In the low-frequency range
the conductivity is similar to the one obtained from the six- and
eight-band models, as expected. However, the additional transition
between the valence band and the antisite band appears around
$800\,\mathrm{meV}$ for the strongly compensated sample
($p=0.025\,\mathrm{nm}^{-3}$ for a Mn concentration of $x=4$\%).
The peak from this transition shifts in the direction of higher
energy with increasing carrier concentration. In the inset we plot
the position of the maximum due to the transition between valence
and defect band as a function of frequency. This dependence can be
useful for rough estimates of carrier concentrations on the basis
of optical and infrared conductivities.

\begin{figure}[tbh]
\includegraphics[width=3.3in,clip]{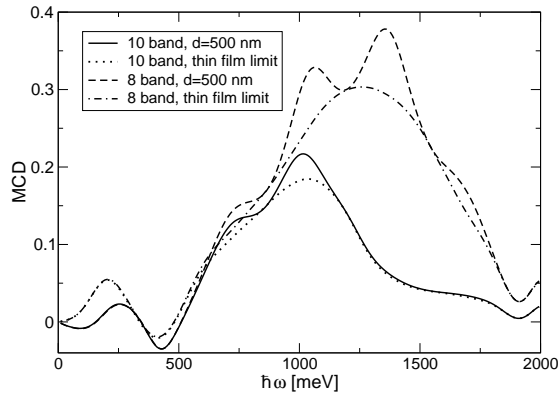}
\caption{Comparison of the theoretical MCD signal for eight- and ten-band
models for a Ga$_{0.95}$Mn$_{0.05}$As epilayer on GaAs substrate.
The dotted and dash-dotted curves correspond to a thin-film geometry.
The solid and long-dashed curves refer to
a film of thickness $d=500\,\mathrm{nm}$.}
\label{fig7}
\end{figure}

The MCD signal should change with the thickness of the (Ga,Mn)As film
(geometry effect) as well as with the number of allowed transitions
(band-structure effect). In the thin-film geometry the condition
$d/\lambda<1$ should be fulfilled, leading to $d<200\,\mathrm{nm}$ for
visible light. However, the (Ga,Mn)As films measured so far have a
thickness of 100--500\,nm. The results presented below consider the
structure air/(Ga,Mn)As/GaAs and take multiple reflections into account.

In Fig.~\ref{fig7} we compare the theoretical results for the MCD signal
for eight- and ten-band models in the thin-film limit and for a film with a
larger thickness of $d=500\,\mathrm{nm}$. Note that the interference peaks
appear for maximal values of $e^{id/\lambda}$ and are superimposed on the
MCD signal for $d=500\,\mathrm{nm}$. The calculations for different film
thicknesses show that the interference peaks change positions, as expected.
However, the common effect is an increase of the MCD signal around the
optical transitions. The difference of theoretical predictions for eight-
and ten-band models is quite pronounced.
The maximum at low photon energies corresponds to transitions
between valence subbands, as discussed above, and is seen for both eight-
and ten-band models. However, for the eight-band model the amplitude of MCD
at high photon energies is almost twice that for the ten-band model. The
main MCD signal is concentrated around the valence-conduction-band
transition for the eight-band model while it is around the
valence-defect-band transition for the ten-band calculation.
%We believe that this is a $f$-sum rule result which gives
%a much larger spectral weight for localized states.
These results agree qualitatively with the experiments.
The MCD spectra of LT-GaAs \cite{Meyer84} show that the MCD signal for
antisites lies around $1\,\mathrm{eV}$ in agreement with the ten-band
model.
The theoretical MCD spectrum has a much larger amplitude for the eight-band
model and is shifted to higher energies in agreement with
experiments.\cite{Beschoten:1999_}

\section{Summary}
\label{summary}

We have presented a detailed analysis of the optical and infrared
conductivity as well as of magneto-optical effects in the infrared
to optical range. The  mean-field/VCA approach using the eight-
and ten-band KL model coupled  with $S=5/2$ Mn spins by a local
exchange interaction has been applied. We have shown that the
ten-band KL model with additional 
disspersionless bands simulating phenomenologically the upper-mid-gap states
is successful in explaining qualitatively the recent infrared and otical 
conductivity measurements by Singley \textit{et al.}\cite{Singley:2003_}
Also, we have predicted strong novel signatures in the magneto-optical
effects in the mid-infrared range which are due to the strong spin-orbit coupling
present in these materilas which allow inter-valence band transitions in
these highly doped samples.

We hope that the predicted dependence of the optical conductivity as well
as the MCD signal and the Kerr effect on the concentrations of carriers
will motivate new experimental efforts to  understand the fascinating
magneto-optical properties of diluted magnetic semiconductors. In
particular, this dependence can provide a test of the validity of the
effective Hamiltonian model in describing the optical properties of these
materials.

\section*{ACKNOWLEDGEMENTS}
We would like to thank W. A. Atkinson, A. H. MacDonald, H. Ohno, K. S.  Burch, D. N. Basov, 
and J. Ma\v{s}ek, for useful discussions.  The work by TD was partly supported by FENIKS
project (EC:G5RD-CT-2001-00535). This work was further supported by the Welch Foundation,
the DOE under grant DE-FG03-02ER45958, and the Grant Agency of the Czech Republic
under grant 202/02/0912.

\end{document}